\documentclass{aastex61}
\usepackage{amsmath}
\accepted{for publication in the {\it Journal of Space Weather and Space Climate}}

\shorttitle{Density structure in sheath}
\shortauthors{Kwon \& Vourlidas}

\begin{document}

\title{THE DENSITY COMPRESSION RATIO OF SHOCK FRONTS ASSOCIATED WITH CORONAL MASS
EJECTIONS}

\correspondingauthor{Ryun-Young Kwon}
\email{rkwon@gmu.edu}

\author[0000-0002-0786-7307]{Ryun-Young Kwon}
\affil{College of Science, George Mason University, 4400 University Drive, Fairfax, VA 22030, USA}

\author{Angelos Vourlidas}
\affiliation{The Johns Hopkins University Applied Physics Laboratory, Laurel, MD 20723, USA}
\affiliation{Also at IAASARS, National Observatory of Athens, GR-15236, Penteli, Greece}



\begin{abstract}
We present a new method to extract the three-dimensional electron density profile and density compression ratio of shock fronts associated with Coronal Mass Ejections (CMEs) observed in white light coronagraph images. We demonstrate the method with two examples of fast halo CMEs ($\sim$ 2000 km s$^{-1}$) observed on 2011 March 7 and 2014 February 25. Our method uses the ellipsoid model to derive the three-dimensional (3D) geometry and kinematics of the fronts. The density profiles of the sheaths are modeled with double-Gaussian functions with four free parameters and the electrons are distributed within thin shells behind the front. The modeled densities are integrated along the lines of sight to be compared with the observed brightness in COR2-A, and a $\chi^2$ approach is used to obtain the optimal parameters for the Gaussian profiles. The upstream densities are obtained from both the inversion of the brightness in a pre-event image and an empirical model. Then the density ratio and Alfv\'{e}nic Mach number are derived. We find that the density compression peaks around the CME nose, and decreases at larger position angles. The behavior is consistent with a driven shock at the nose and a freely-propagating shock wave at the CME flanks. Interestingly, we find that the supercritical region extends over a large area of the shock and last longer (several tens of minutes) than past reports. It follows that CME shocks are capable of accelerating energetic particles in the corona over extended spatial and temporal scales and are likely responsible for the wide longitudinal distribution of these particles in the inner heliosphere. Our results also demonstrate the power of multi-viewpoint coronagraphic observations and forward modeling in remotely deriving key shock properties in an otherwise inaccessible regime. 
\end{abstract}




\section{INTRODUCTION} \label{sec:intro}
Shocks associated with Coronal Mass Ejections (CMEs) are one of the sources responsible for highly energetic particles, called Solar Energetic Particles \citep[SEPs; e.g., see reviews by][]{R1999,D2016}. While particle acceleration by flares \citep[see][]{R1999} is expected to occur in a limited volume (i.e., magnetic reconnection site), fast-mode coronal shocks are able to directly inject SEPs over a broad range in heliolongitudes \citep[e.g.,][]{Cl1995,Cl2005}. SEPs are an important component of space weather as the cause of radiation hazards to astronauts and satellites, orbital degradation of satellites, communication disruptions, and electrical blackouts \citep[e.g.,][and references therein]{B2014}. To better understand how coronal shocks accelerate particles over a wide range of heliolongitudes and hence improve our ability to predict their intensity, duration and energy spectrum, it is important to understand the properties of coronal shocks associated with CMEs and their temporal and spatial relationship with the SEPs measured in interplanetary space (IP).

In IP, the association between CME-driven shocks and particles is known for a long time thanks to direct in-situ measurements of shocks and SEPs \citep[e.g.,][]{C1988}.
This has been difficult in the corona because direct measurements are currently unavailable, although the situation will soon be remedied \citep{mccomas2016}. Understanding the formation and evolution of shocks in the corona is crucial for discriminating between flare and CME origins, particularly for the highest energy (GeV) SEPs ejected when CMEs are at heliocentric distances of around 2-10 solar radii \citep[$R_\odot$;][]{tylka2005}.

Coronal shocks are observed via remote-sensing observations in Extreme Ultraviolet (EUV), white light, and radio imaging and via spectroscopy \citep{voubempo2012}. They form large-scale spherical fronts seen as halos by white light observations, or as EUV waves propagating against the solar coronal base \citep[e.g.,][see also recent reviews by \citet{Pat2012},\citet{W2015}, and \citet{L2017} for debates on the nature of EUV waves]{K2017}. Occasionally, type II radio bursts over a large spectral range accompany these shocks, particularly in IP space \citep[e.g.,][]{Gopal2008}, while metric Type II emission seems to originate from the flanks of shock waves close to the Sun \citep[e.g.,][]{demoulin2012}. 

There is a growing amount of evidence for the association of CME-driven shocks with SEPs in the corona \citep[see recent papers, e.g.,][and references therein]{R2012,R2016,Ca2013,La2014,La2016,S2016} but it remains unclear whether such shock waves are capable of accelerating particles at the observed energies \citep{B2010,B2011}. An important shock parameter that could be accessible from coronagraphic observations is the ratio between the downstream and upstream electron densities or density compression ratio, $X$. According to the diffusive shock acceleration theory, the slope of the SEP intensity spectrum depends only on the density compression ratio \citep[e.g. Eq. (11) in][]{D2016}. Therefore, measurements of the compression ratio and its temporal evolution in the corona can go a long way in understanding SEP in situ observations in the inner heliosphere.

Because the coronagraph images provide only the projected density, some model or knowledge of the density distribution along the line of sight (LOS) is necessary to obtain an estimate of the true volumetric electron density across the shock. \citet{O2009} were the first to extract estimates of the compression ratio for a number of shock fronts in the Large Angle and Spectrometric Coronagraph \citep[LASCO;][]{B1995} field of view using forward modeling techniques. However, those were single viewpoint observations and hence the shock reconstructions might be subject to considerable uncertainties.

Here, we return to this issue employing multi-viewpoint observations from the Solar Terrestrial Relations Observatory \citep[\textit{STEREO};][]{Ka2008} Sun-Earth Connections Coronal and Heliospheric Investigation \citep[SECCHI;][]{H2008} COR2 coronagraph and more sophisticated forward models and analysis methods. We reconstruct the three-dimensional (3D) electron density at the shock sheath using an ellipsoidal forward model and populating it with a thin shell of Gaussian distributed electron density. The upstream density is derived with two methods: using the \citet{L1998} model and via the inversion of a pre-event polarized brightness image. The method is applied to two CMEs, on 2011 March 7 and 2014 February 25, associated with intense SEP events \citep{P2013,La2016}.

The paper is organized as follows. In Section \ref{sec:method}, we present the model and method used to determine the electron density distributions at the shock sheaths. In Section \ref{sec:results}, we present the resulting sheath (downstream) and upstream electron densities and density ratios and infer  Alfv\'{e}nic Mach numbers. Our summary and conclusion are given in Section \ref{sec:con}.

   \begin{figure}
   \centering
\includegraphics[width=1\textwidth]{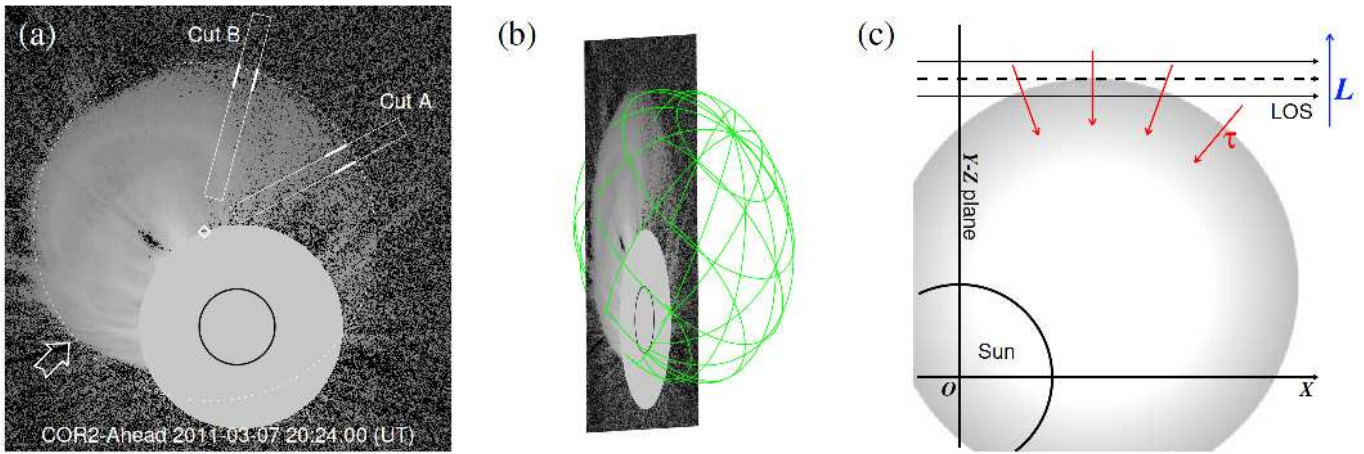}
   \caption{(a) An excess brightness COR2-A image taken at 2011 March 7 20:24 (UT). The projection of the 3D shock front (green lines in panel (b)) on the image plane is shown with the dotted line. The diamond refers to the geometric center of the ellipsoid model projected onto the same plane. Two examples of radial cuts (relative to the geometric center), used to take the brightness profile, are shown by the two rectangles marked as Cut A and Cut B. Thick lines in the rectangles show the length of $L$ used for the calculation of $\chi^2$ (see Figure \ref{fig:bp}). (b) The excess brightness in panel (a) with the 3D shock front (green lines). The 3D geometry was modeled with the ellipsoid model in \citet{K2017}. (c) The geometric relation among the Sun, shell-like sheath, and LOS. A partial circle around the origin $O$ is the solar disk. A shell-like sheath is represented in gray color. Arrows in black, blue and red are the LOS, the projected shock normal on the image plane, and the actual shock normal in 3D, respectively.} \label{fig:illust}
   \end{figure}
      
\section{METHOD} \label{sec:method}
\subsection{Data}
The 3D reconstruction of these CMEs and shocks has been reported in \citet{K2017} using the Graduated Cylindrical Shell \citep[GCS;][]{Tn2006} model and the ellipsoid model model \citep{K2014}, respectively. The fits were based on three viewpoint observations from STEREO-A, -B and SDO \citep[Solar Dynamics Observatory;][]{P2012}/SOHO \citep[SOlar and Heliospheric Observatory;][]{D1995}. For the density ratio analysis, we use the COR2-A images because of their higher signal-to-noise ratio. The COR2 white light images capture Thomson-scattered photospheric light by the electrons in the corona in a field of view (FOV) of 2.5-15 $R_\odot$.  We do not analyze any COR1 images for these events because CME-associated shock waves are generally too faint in these heights to allow extraction of profiles.
   
\subsection{3D Model of Density Structure in Sheaths} \label{subsec:sd}
Our density analysis method uses the 3D geometry of shocks. Figure \ref{fig:illust} shows the geometric relation among the 2D image plane, the spherical shock front in 3D and the LOS. Panel (a) shows an excess brightness due to the CME and shock, for example, taken at 2011 March 7 20:24 (UT) by COR2-A. The halo front on the image plane (dotted line in panel (a)) is the projection of a bubble-shaped shock front (green lines in panel (b)). The 3D geometry of the bubble-shaped shock fronts is modeled with the ellipsoid model \citep{K2014}. The sheath structure is given as a shell behind the ellipsoid model. Panel (c) shows a cross-section of the shell-like sheath. The image plane of observations (panel (a)) is defined as the $Y$-$Z$ plane, and the $X$-axis is towards the observer. The origin $O$ is at Sun center. $L$ with an arrow in blue color is the distance normal to the projected shock front on the image plane. $L$ is measured from the edge of the shock front on the image plane (dashed LOS arrow in this panel). Note that the shock normal on the image plane differs from the actual shock normal in 3D. The shock normal in 3D is shown by arrows in red, and $\tau$ denotes the distance normal to the shock front in 3D from the edge of the shock where the electron density due to the shock begins increasing. The excess brightness on the image plane results from the integration over the LOS passing through the shell-like sheath.

\begin{figure}
\begin{center}
\includegraphics[width=1\textwidth]{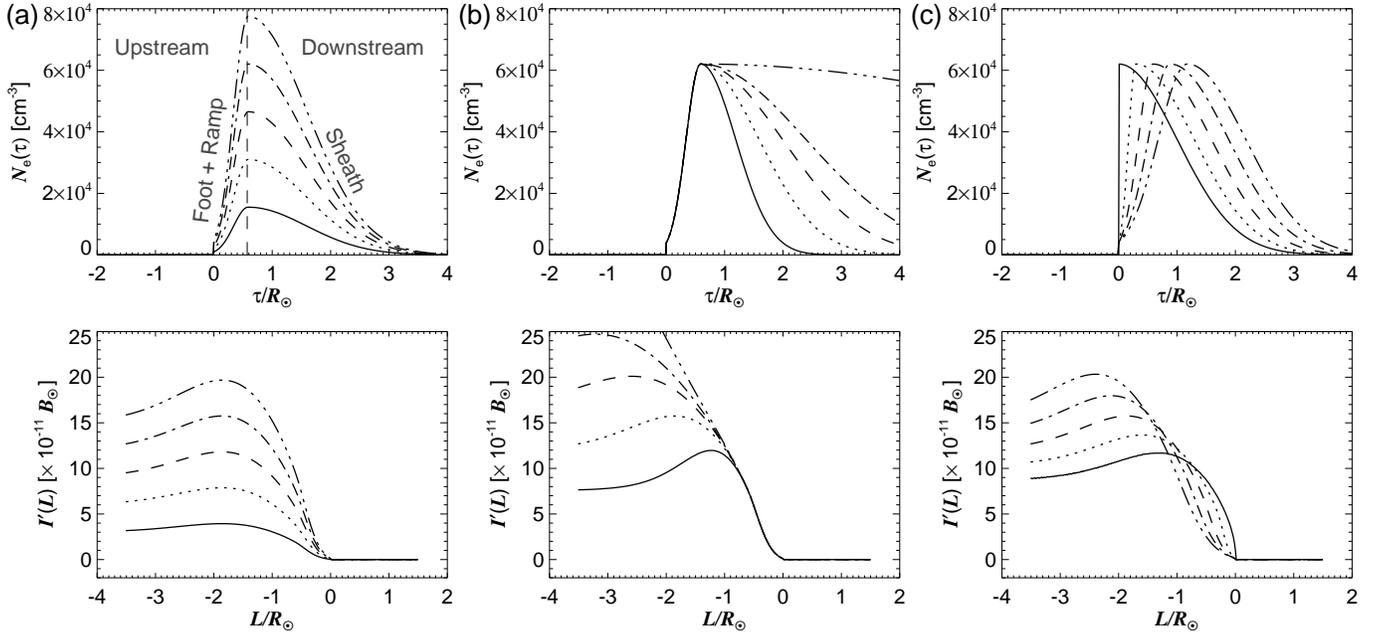}
\caption{The upper three panels show the dependency of the sheath electron density profiles ($N_e(\tau)$) on the three parameters, $\rho_e$ (panel (a)), $d_\text{sheath}$ (b), and $d_\text{front}$ (c), while the other two parameters are fixed. The lower panels are the modeled Thomson-scattering brightness profiles $I^\prime(L)$ in mean solar brightness ($B_\odot$), resulting from the given $N_e(\tau)$ in the same line-styles in the upper panels. Lines in solid, dotted, dashed, dash-dotted, dash-dot-dotted refers to the cases when $\rho_e$ =  (1.6, 3.1, 4.7, 6.2, 7.8) $\times$ 10$^4$, $d_\text{sheath}$ = 1\,$R_\odot$, and $d_\text{front}$ = 0.25\,$R_\odot$ in panel (a); $\rho_e$ =  6.2 $\times$ 10$^4$, $d_\text{sheath}$ = (0.6, 1.0, 1.4, 1.8, 8.0)\,$R_\odot$, and $d_\text{front}$ = 0.25$R_\odot$ in panel (b); and $\rho_e$ =  6.2 $\times$ 10$^4$, $d_\text{sheath}$ = 1\,$R_\odot$, and (0.0, 0.1, 0.3, 0.4, 0.5)\,$R_\odot$ in panel (c). Because the Gaussian tail is not zero even at large distances, $N_e(\tau < 0)$ is not exactly zero, if $d_{\text{front}}$ $\neq$ 0.  By definition, we simply set $N_e(\tau < 0)$ = 0. The 3D geometry of the shock used to calculate $I^\prime$ is shown in Figure \ref{fig:illust}(b).} \label{fig:sm}
\end{center}
\end{figure}

We would like to emphasize that the electron density jump along the 3D shock normal $\tau$ cannot be determined directly from the brightness profile along the projected shock normal $L$. As shown in Figure \ref{fig:illust}(c), a LOS passes through various $\tau$, and the observed excess brightness is the integration over the LOS. A way to overcome this is to model the excess brightness $I^\prime(L)$ integrating the sheath electron density ($N_e(\tau)$) along the LOS ($s$) and compare it with the observed excess brightness $I(L)$. Once we obtain the best $I^\prime(L)$ for $I(L)$, $N_e(\tau)$ along the shock normal is determined. To model $N_e(\tau)$, we use Gaussian function that is generally used for the density structure of a wave, but allow asymmetry by using two Gaussian functions, namely,
\begin{eqnarray}
N_e(\tau)=\rho_e\exp{-\frac{(\tau-\tau^\prime)^2}{2d^2}} \, , \\ \label{eq:ne_s}
d=\begin{cases}
d_{\text{front}}, \, \text{if}\,  \, \tau < \tau^\prime\, , \\ \nonumber
d_{\text{sheath}}, \, \text{if}\,  \, \tau \geq \tau^\prime\, ,
\end{cases} \\
\text{with}\, \,\tau^\prime=2d_{\text{front}}\sqrt{2\ln{2}} \, , \nonumber
\end{eqnarray}
where $\rho_e$ and $d$ are constants. $\tau^\prime$ is the full width at half maximum of the Gaussian function with $d_{\text{front}}$. The outermost edge of the density structure is defined at $L$ = 0, so that $N_e(\tau < 0)$ = 0. $\rho_e$ is the peak electron density excess (density jump).

The upper panels in Figure \ref{fig:sm} show the resulting sheath electron density distributions $N_e(\tau)$ from various sets $\rho_e$, $d_{\text{front}}$, and $d_{\text{sheath}}$. For instance, panel (a) shows $N_e(\tau)$ obtained when varying $\rho_e$ while the other two parameters are fixed. In general, $N_e(\tau)$ starts increasing at $\tau$ = 0, reaches its maximum at $\tau$ = $\tau^\prime$, and decreases gradually where $\tau$ $>$ $\tau^\prime$. The key features of the function and the physical interpretations can be summarized as follows.
\begin{enumerate}
\item If $d_\text{front}$ = $d_\text{sheath}$, it becomes a linear wave.
\item If $d_\text{sheath}$ is very large (dash-dot-dotted line in panel (b)) and $d_\text{front}$ = 0 (solid line in panel (c)), $N_e(\tau)$ becomes a step function near the shock front, followed by a gradually decreasing tail. This profile can represent steepening waves and shock waves.
\item if $d_\text{front}$ $\neq$ 0 and $d_\text{front}$ $<$ $d_\text{sheath}$ (panel (a)), it resembles the in-situ measurements and the theoretical expectations of collisionless shocks \citep[see a review by][]{Tr2009}. In this case, $d_\text{front}$ serves as the density in the foot and ramp, with $d_\text{sheath}$ being the density in the sheath including overshoot and turbulence (if any).
\end{enumerate}
The real density distribution may deviate from our double-Gaussian model. Note, however, that the total number of electrons will largely depend on the density excess $\rho_e$, rather than the detailed shape of the function or fine scale structures (e.g., foot and overshoot). Since we fit the density model to observed brightnesses that result from the LOS integrations, we expect that the details of the profile is negligible when determining the density jump at shocks (see Figure \ref{fig:ex1}). Note that Equation (\ref{eq:ne_s}) is similar to the function in \citet{Tn2006}. \citet{Tn2006} used the function to reproduce the enhanced brightness due to the pileup plasma of an expanding flux rope, whereas the function is used for the density structure of waves and shocks in this paper.

Since observations do not resolve density distributions along the shock normal as discussed above, we calculate the Thomson-scattering brightness corresponding to the given density distribution $N_e(\tau)$ to be compared with the actual observation. We use the ellipsoid fits in \citet{K2017} as the 3D geometry of the shock fronts. Given the 3D geometry as illustrated in Figure \ref{fig:illust}(c), the 3D coordinates of all points along the LOS are known, and thus $\tau$ along the LOS is also known.

Once the sheath electron distribution $N_e(\tau)$ and the coordinates of the LOS are given, Thomson-scattering brightnesses resulting from the integration can be calculated using the standard Thomson-scattering equations \citep[e.g.][]{B1966}. We assume that the sheath electron density structure varies slowly along the shock surface so that a single density distribution model can represent the density structure for the region where the LOS are passing through (see Figure \ref{fig:illust}(c)). This assumption is valid because the fits of the model to observations are done only close to the shock front (see Section~\ref{sec:de}).

The lower panels of Figure \ref{fig:sm} show the calculated Thomson-scattering brightnesses $I^\prime(L)$ from electron density distributions $N_e(\tau)$ given in the upper panels, together with the 3D geometry of the shock shown in Figure \ref{fig:illust}(b). Note that the shapes of $I^\prime(L)$ differ from the sheath density distributions $N_e(\tau)$. For instance, $I^\prime(L = 0)$ at the shock front is zero, although the electron density has the maximum at the shock front (see the solid line in Figure \ref{fig:sm}(c)). These differences are due to the integration along the LOS.

As shown in Figure \ref{fig:sm}, the shape of $I^\prime$ varies with the density model $N_e(\tau)$ for the given geometries. In this sense, we find the optimal parameters of $N_e(\tau)$ by minimizing $\chi^2$ defined as
\begin{equation}
\chi^2=\frac{1}{m}\left(\frac{I(L^\prime-\Delta L)/I^\prime(L)-1}{w}\right)^2 \, , \label{eq:chi2}
\end{equation}
where $I$ and $I^\prime$ are the observed and modeled excess brightnesses, respectively. $m$ is the total number of the points compared, and $w$ is the weight function. $\Delta L$ is the displacement on the sky plane of the true shock front from the initially estimated one, i.e., $L$ = $L^\prime$ - $\Delta L$. Because of this, our fit has the four parameters, $\rho_e$, $d_{\text{front}}$, $d_{\text{sheath}}$, and $\Delta L$. Note that the difference between $I$ and $I^\prime$ is normalized by $I^\prime$. Instead of the error in $I$, we use a weight function $w$ for the denominator, and the weight function is defined as,
\begin{equation}
w=\left[1+\left(\frac{|L|}{R_\odot}\right)^{1/3}\right]^{-1} \, .
\end{equation}

\subsection{Upstream Density Profile} \label{subsec:up}
We estimate the upstream electron density $\rho_u$ via inversion of a pre-event polarized brightness image.  The method assumes an axisymmetric density distribution that can be described with an $n$th-order polynomial function \citep{van1950,hayes2001}. We use the polynomial form in \citet{L1998},
\begin{equation}
\rho_u(r)=A\left(\frac{r}{R_\odot}\right)^{-2}+B\left(\frac{r}{R_\odot}\right)^{-4}+C\left(\frac{r}{R_\odot}\right)^{-6} \, , \label{eq:up_ne}
\end{equation}
where $A$, $B$, and $C$ are constants. The constants are found by the fit to the observed profiles at each position angle. Equation (\ref{eq:up_ne}) is essentially equivalent to the standard polynomial expansion used traditionally in the field. Note that it reduces the number of constants while it includes the higher-order terms. 

Because of the presence of F-corona polarized component, especially above 5 Rs \citep{hayes2001}, the determined $\rho_u$ in our height range would be overestimated. In contrast, the measured $\rho_u$ in coronal holes could be {\it suppressed} because of the calibration \citep[e.g.,][]{hayes2001,T2010}. We have checked other empirical coronal electron density models that have been used widely, i.e., \citet{N1961}, \citet{S1977}, and \citet{L1998}, to infer $\rho_u$. Since the \citet{L1998} model provides the lowest electron density among them, we use as a lower limit. In this way, the upstream electron density $\rho_u$ is obtained as a range between our measurement and the \citet{L1998} model.

\begin{figure}
\begin{center}
\includegraphics[width=.9\textwidth]{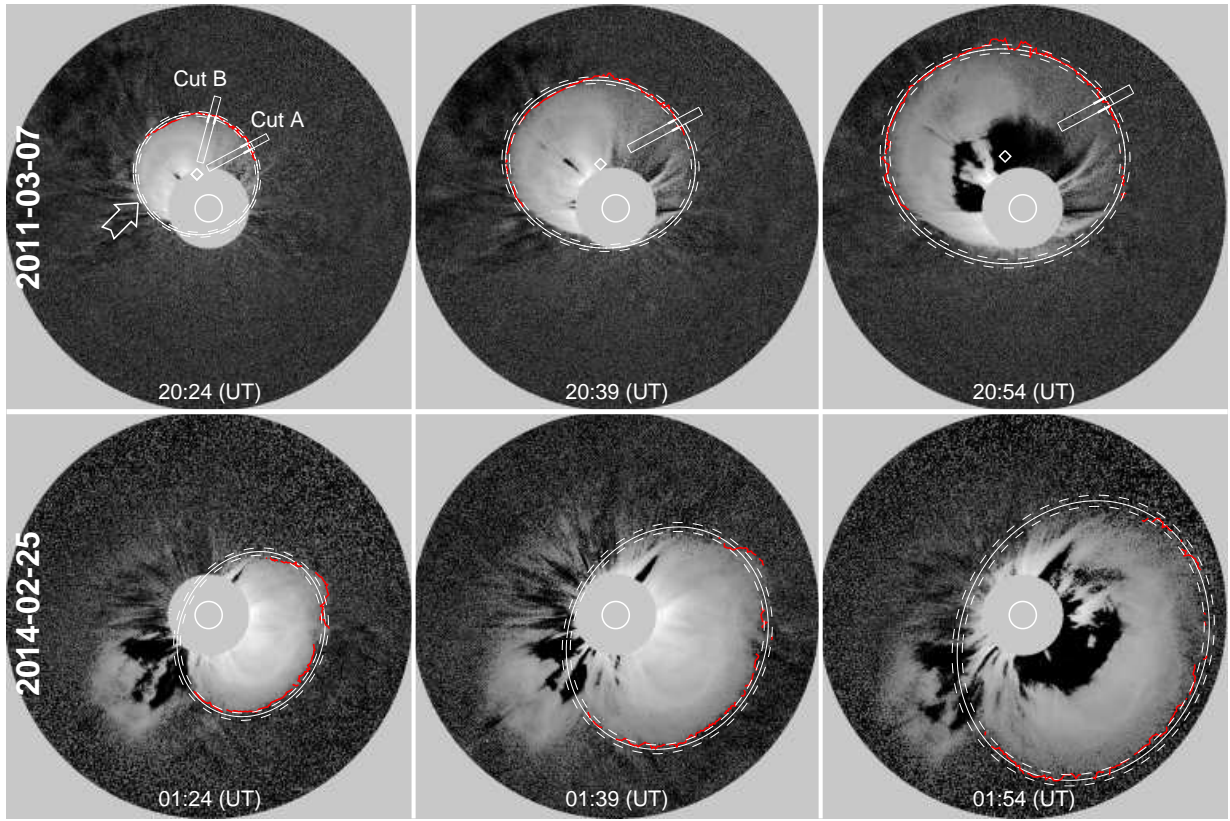}
\caption{Excess brightness images used to determine the density excess $\rho_e$. The upper and lower panels show two different events on 2011 March 7 and 2014 February 25, respectively. Dots in red are the points where the $\rho_e$ measurements have been made. Solid lines refer to the projected shock front determined with the ellipsoid model. Dashed lines represent the $\pm$ 4\% error in the geometry of the shock fronts.} \label{fig:obs}
\end{center}
\end{figure} 

\subsection{Error in Electron Density Excess} \label{sec:err}
The main sources of error in $\rho_e$ are the uncertainties of brightness given by white light observations, the LOS integration relying on the ellipsoid model, the Gaussian description of the density distribution, and the relation between the white light structure and the true shock.

First of all, we use the excess brightness to determine $\rho_e$. An excess brightness image is obtained by subtracting a previous image, which is temporally closest but at least 30 minutes apart from the image. In this manner, it is expected that the noise level, including the calibration error, the F-corona polarized component, the stray light component, and the brightness due to the background electron density, is similar in the two images, since they are temporally close. In this sense, the error in the excess brightness would be less than the error \citep[20\%;][]{frazin2012} of the direct brightness images. In addition, the CME and shock structure in the previous image cannot affect the excess brightness close to the shock front, because the two images are at least 30 minutes apart. Note that the shocks/waves in white light observations are generally faint and diffuse \citep[e.g.,][]{S2000}, so that the signal-to-nose level is still a issue. We use the COR2-A images because of their higher signal-to-noise ratio than the COR2-B images.

The 3D geometry is the key to deriving electron densities from white light observations. We rely on a highly idealized ellipsoid description for the 3D geometry of the shock front, and thus the LOS integration. Obviously the real shock front may deviate from that. Multi-spacecraft in-situ measurements at 1 au have shown that there exist the shock surface ripples so that the shock surface cannot be represented by a planar or spherical structure \citep[][and references therein]{N2005}. We believe that this effect is negligible in the corona. Instead, we determine the error in the ellipsoid model as shown in the Appendix in \citet{K2014}. This error gives us the difference of the ellipsoid model from the observed front. Because of the error ($\pm$ $\delta$) in the 3D geometry, we have the three ellipsoid models (see dashed lines in Figure \ref{fig:obs}), and the lower and upper estimates of the 3D geometry are used to estimate the error in $\rho_e$.

The sheath electron density distribution model in Equation (\ref{eq:ne_s}) is also highly idealized, and the real density distribution may deviate from our model. However, the brightness is obtained by taking the integration of the density over the LOS. It is obvious that the small scale structure will not affect much the total number of electrons and, therefore, the observed brightness. As we will see in Section in \ref{sec:de}, the most significant source causing the brightness increase is the peak electron density excess $\rho_e$ (density jump).

The imperfection of the background subtractions will result in error in excess brightness. If we assume that the true $\rho_e$ varies slowly along the shock surface, the error can be estimated by comparing $\rho_e$ with those in the neighboring position angles. We repeat the same analysis for every position angles and take the average and standard deviation over 7$^\circ$. The standard deviation serves as the error. Note that we discard the cases where the brightness profiles are contaminated by the following pileup plasma or deflected streamers.

As discussed above, the two errors in $\rho_e$ are given from the uncertainties of the ellipsoid model and the excess brightness. We use the larger one.

\section{RESULTS AND DISCUSSION} \label{sec:results}
We apply our method to two fast CMEs on 2011 March 7 and 2014 February 25. The details of the 3D reconstructions with the ellipsoid model are given in \citet{K2017} but we summarize the modeling results here for completeness. The shock speeds in 3D are $\sim$ 2200 km s$^{-1}$ and $\sim$ 2050 km s$^{-1}$ for the March and February events, respectively. The minimum angular widths of the shock envelopes are 192$^\circ$ and 252$^\circ$, whereas the CME angular widths are 58$^\circ$ and 90$^\circ$, respectively. The shock speeds vary with position angle. The maximum speeds are seen near the CME noses, in which the speeds are well correlated with the CME nose speeds. The speeds in the far-flanks tend towards the local fast magnetosonic speed.

This 3D modeling considered only the outline of the CME and its outer shock envelope. Here, we take the next step and attempt to fit the observed brightness of the shock by modeling the shock envelope as a thin shell with some electron density distribution. Thus we can determine the density excess $\rho_e$ at the sheaths and the downstream--upstream density ratios $X$, using the 3D geometry of the shocks. The 3D geometries and kinematics and the estimated electron densities enable us to also estimate the Mach numbers $M_A$ and upstream Alfv\'{e}n speeds $v_A$.

\begin{figure}
\begin{center}
\includegraphics[width=.5\textwidth]{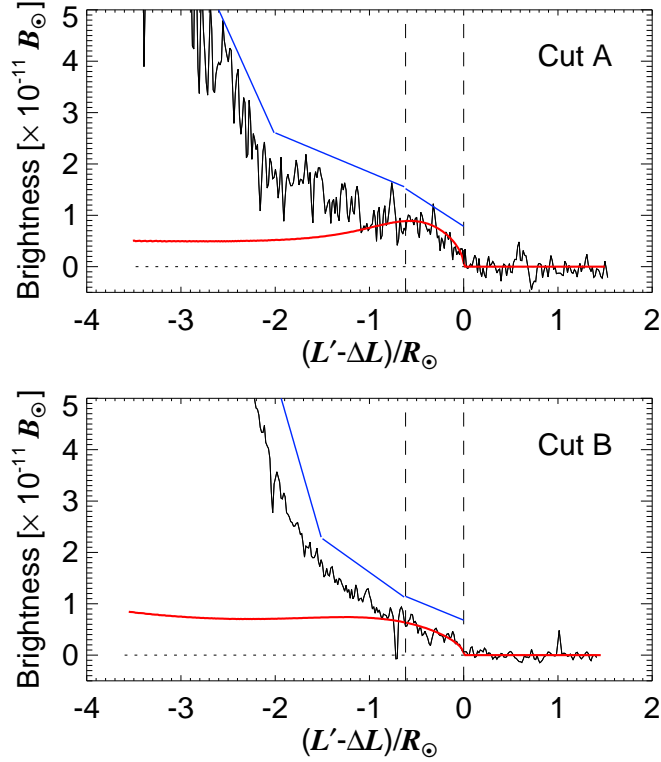}
\caption{Excess brightness profiles $I(L=L^\prime - \Delta L)$ taken from Cut A (upper panel) and Cut B (lower) shown in Figure \ref{fig:obs}. The dotted line in each panel indicates the zero brightness. The red line in each panel is the modeled excess brightness profile $I^\prime(L)$ obtained from $\chi^2$ fits shown in Figure \ref{fig:ex1}. The blue lines indicate the slope of the observed brightness profiles. The two dashed lines in each panel demarcate the region where the $\chi^2$ values are calculated.} \label{fig:bp}
\end{center}
\end{figure}

\subsection{Density Excess $\rho_e$ at Shock Front} \label{sec:de}
Figure \ref{fig:obs} shows the excess brightness COR2-A images for two events on 2011 March 7 (upper panels) and 2014 February 25 (lower panels). The previous images (at least 30 minutes apart) are subtracted from these images, so the brightness is largely due to the electrons in the CME and sheath. To derive the density structure of the sheath we take radial cuts (relative to the geometric center of the shock front; see the diamond symbols in the first panel of Figure \ref{fig:obs}) along all position angles at 1$^\circ$ intervals. The position angle, $\zeta$, is measured counterclockwise from the shock leading edge (directional axis of the ellipsoid model projected onto the image plane). The two rectangles in the first panel of Figure \ref{fig:obs} (see also Figure \ref{fig:illust}(a) zoomed in on the shock front) are two example radial cuts, far from and close to the CME nose (Cut A and Cut B, respectively). To increase the signal-to-nose ratio, we average the brightness profiles across the width of the rectangle. The width $\omega$ is chosen as, $\omega$ = $2e_r\sin{\theta}$, where $e_r$ is the average distance of the projected shock front from its geometric center, and $\theta$ = $\cos^{-1}{(1-\lambda/e_r)}$. $\lambda$ is the distance corresponding to 0.5 pixel. In this way, the difference in distance $L$ of the  \textit{curved} shock front across the width is at most 0.5 pixel, and thus the curvature is negligible when taking the average over the width.

\begin{figure}
\begin{center}
\includegraphics[width=1\textwidth]{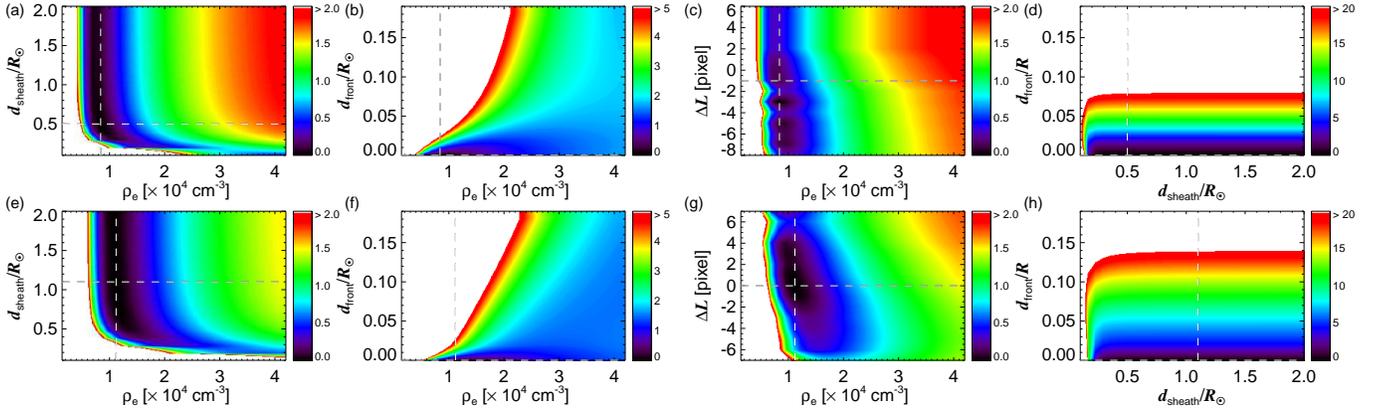}
\caption{Fits of the sheath density distribution model to the observed brightness profiles of Cut A (upper) and Cut B (lower). From the leftmost to the rightmost, the panels show the $\chi^2$ values varying with the parameters, $d_\text{sheath}$--$\rho_e$, $d_\text{front}$--$\rho_e$, $\Delta L$--$\rho_e$ and $d_\text{front}$--$d_\text{sheath}$, respectively, and the $\chi^2$ values are indicated by the color bars. Dashed lines in each panel indicate the parameters at the minimum $\chi^2$.
}  \label{fig:ex1}
\end{center}
\end{figure}

The upper and lower panels of Figure \ref{fig:bp} show the excess brightness profiles of Cut A and Cut B, respectively. Since the background brightness has been subtracted, the profiles are flattened, and the brightness in the region where $L$(= $L^\prime$ $-$ $\Delta L$) $>$ 0 is 0, as indicated by the horizontal dotted line in each panel. While our density model in Equation (\ref{eq:ne_s}) is only for the excess brightness due to the shock, the brightness profiles also contain the brightnesses owing to the plasma pileup ahead of the following CME and deflected streamers.

\begin{figure}
\begin{center}
\includegraphics[width=.5\textwidth]{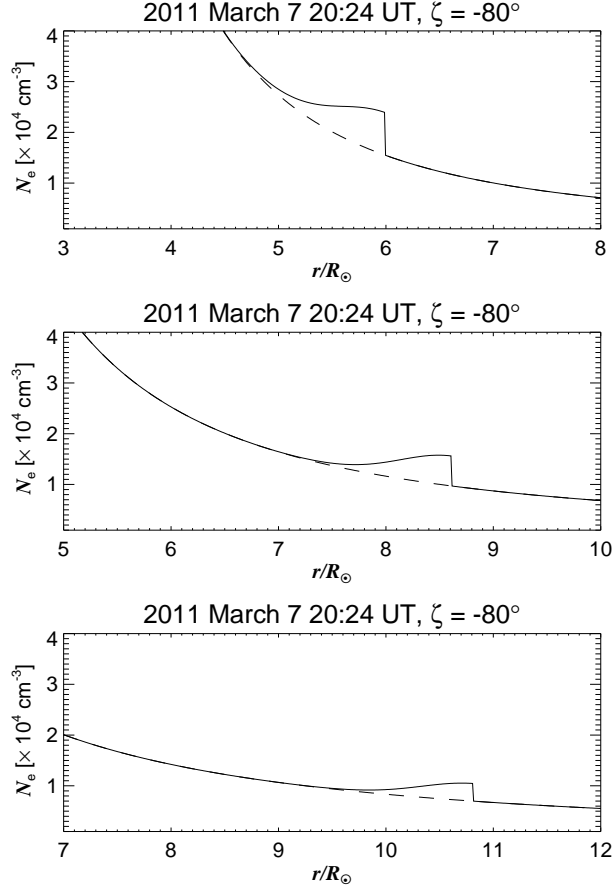}
\caption{Evolution of the sheath taken at Cut A ($\zeta$ = 80$^\circ$) of the 2011 March 7 event (rectangles in the top panels of Figure \ref{fig:obs}). From the top to bottom, each panel shows the modeled sheath electron density distribution plus the background electron density at 20:24 UT, 20:39 UT, and 20:54 UT, respectively. Note that the ranges of the abscissas are different.} \label{fig:evol}
\end{center}
\end{figure}

\begin{figure}
\begin{center}
\includegraphics[width=1\textwidth]{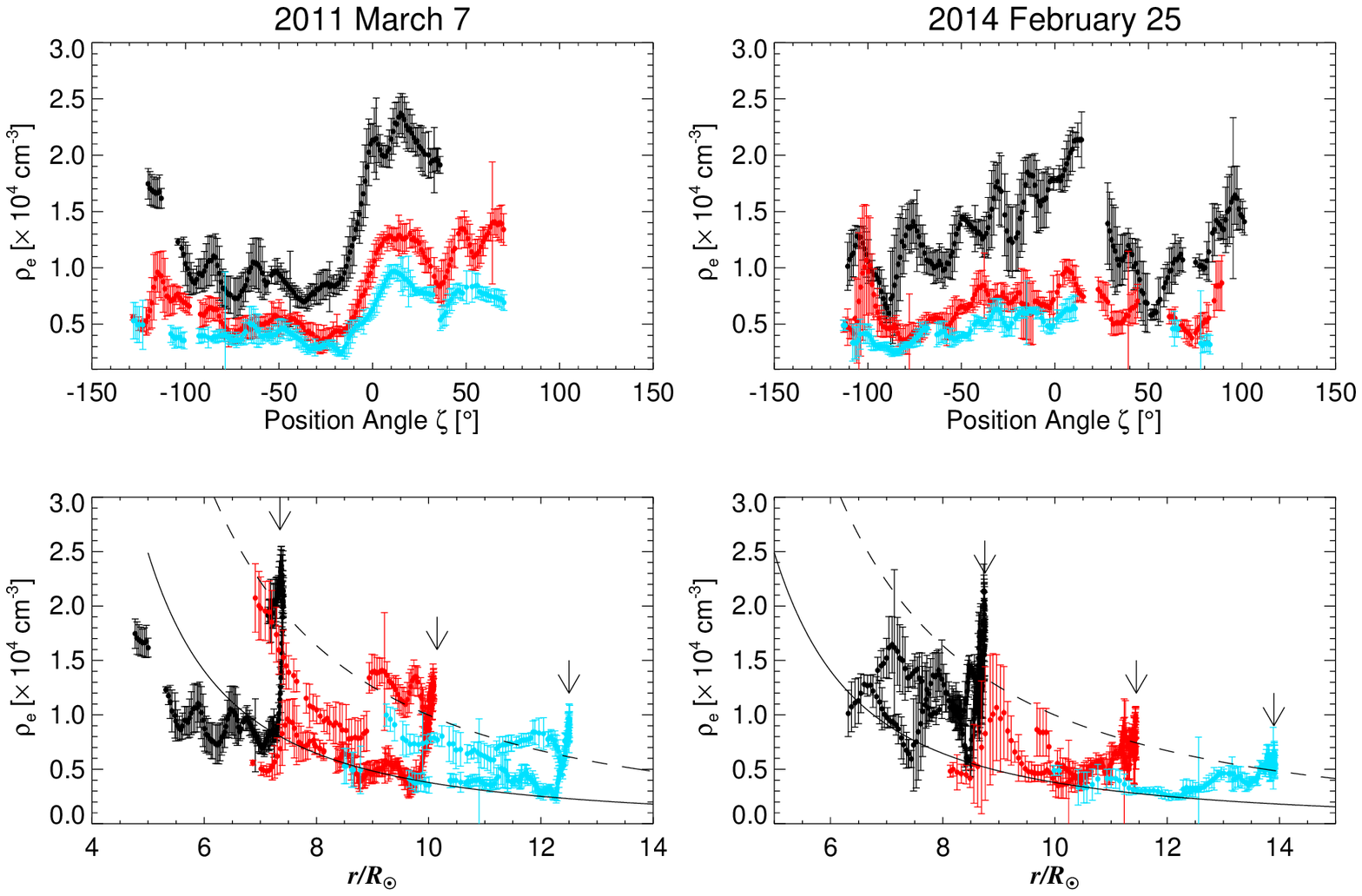}
\caption{Determined electron density excess $\rho_e$ for the 2011 March 7 and 2014 February 25 events. The results were obtained at three different times for each event, and they are shown in different colors, black, red, and cyan in chronological order (20:24 UT, 20:39 UT, and 20:54 UT for the 2011 March 7 event and 01:24 UT, 01:39 UT, and 01:54 UT for the 2014 February 25 event). The upper row shows the $\rho_e$ distributions over position angle $\zeta$. The position angle that $\zeta$ = 0$^\circ$ is at the projected directional axis of the ellipsoid model. The lower row shows the $\rho_e$ distributions over height. Lines in solid and dashed represent two empirical electron density models in \citet{L1998} and \citet{S1977}, respectively.} \label{fig:den}
\end{center}
\end{figure}

In order to discriminate the shock part from the following pileup plasma in the excess brightness profiles, we use the slope of the profiles. The brightness at the outermost part ($L$ $\sim$ 0) is expected to be mostly due to the density jump, but the following pileup plasma enhances the brightness. The top panel in Figure \ref{fig:bp} shows the case that the brightness profile is taken at a position angle far away from the CME. While the brightness due to the pure sheath (red line; see Figure \ref{fig:ex1} and the description) is expected to decrease at large distances ($|L|$), the observed brightness increases with distance. Since the CME is far away from the shock front at this position angle, the effect of the pileup plasma on the brightness profile would be gentle. As indicated by the three blue lines in this panel, it results in the slower increase in brightness in the intermediate part (second blue line; -2 $\leq$ $L$ $\leq$ -0.6) than the outermost part (third blue line; 0 $\leq$ $L$ $\leq$ -0.6). Since the innermost part (first blue line) is mostly due to the pileup plasma, the gentle slope drops off into a steep slant as the distance increases. When the CME is close to the shock front, as seen in the lower panel of Figure \ref{fig:bp}, the slope keeps increasing with distance. The three-part slope enables us to discriminate the part due to the shock and minimizes the effect of the following pileup plasma on the fits. We only use the first increase part as demarcated by the two dashed lines in each panel (see also the thick lines in the rectangles in Figure \ref{fig:obs}). The red dots in Figure \ref{fig:obs} refer to the position angles where we have been able to determine $\rho_e$ using $\chi^2$ minimizations. In the position angles where the following CME is too close, we simply discard them.

\begin{figure}
\begin{center}
\includegraphics[width=.5\textwidth]{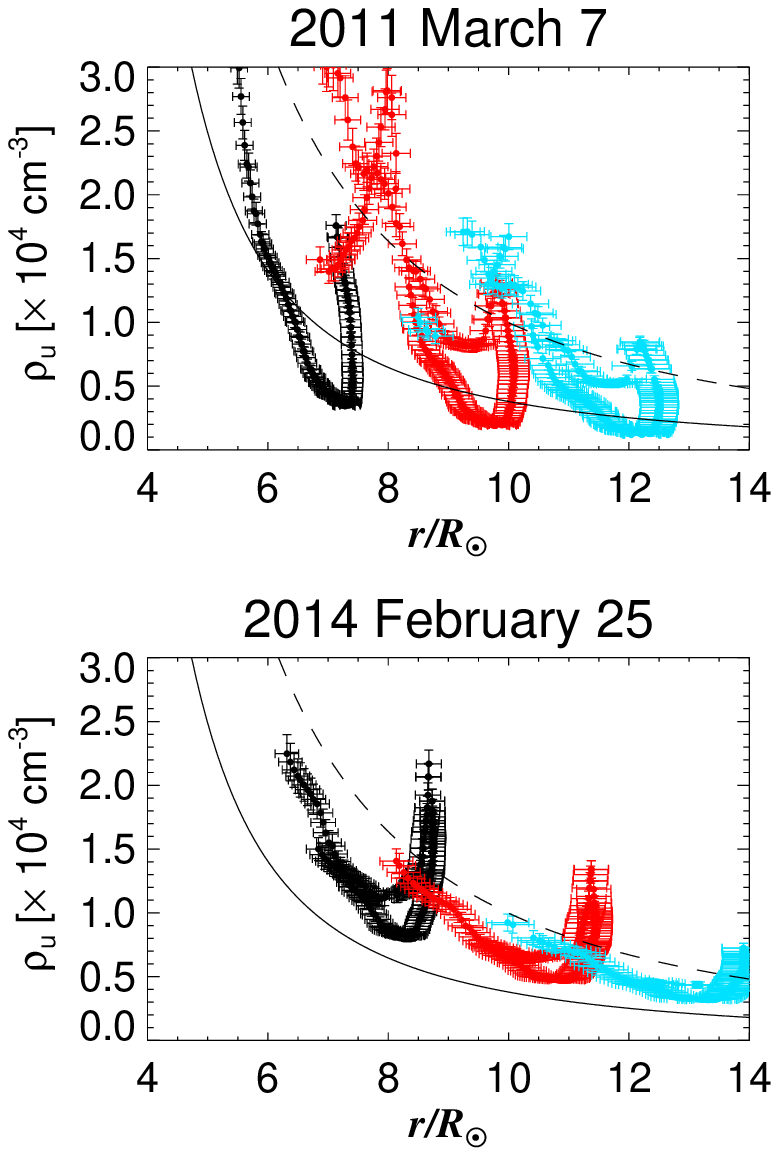}
\caption{Derived upstream electron densities $\rho_u$ for the 2011 March 7 and 2014 February 25 events. The densities determined at three different frons are shown in black, red, and cyan, in accordance with Figure \ref{fig:den}. Lines in solid and dashed represent two empirical electron density models in \citet{L1998} and \citet{S1977}, respectively. } \label{fig:up}
\end{center}
\end{figure}

\begin{figure}
\begin{center}
\includegraphics[width=.9\textwidth]{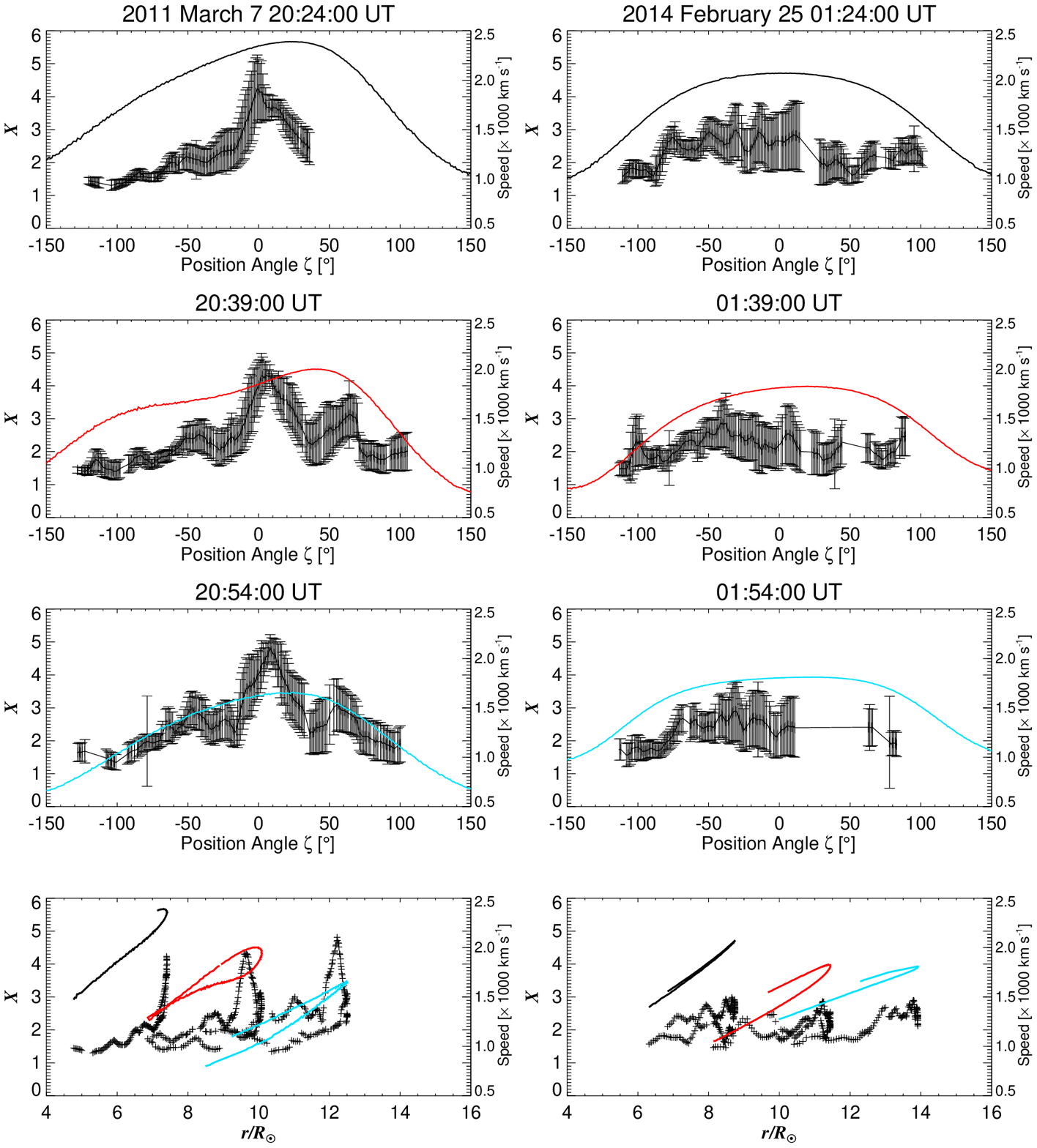}
\caption{Density ratios of downstream over upstream $X$ = $\rho_d$/$\rho_u$. The left and right columns are the results for the 2011 March 7 and 2014 February 25 events, respectively. The three rows from the top show $X$ over position angle $\zeta$. The bottom panels show the determined $X$ over height. Solid lines in black, red and cyan are the speeds of the shock fronts derived from the 3D ellipsoid model, in accordance with the ordinate on the right-handed side.} \label{fig:X}
\end{center}
\end{figure}

Figure \ref{fig:ex1} describes our $\chi^2$ approach to obtain the optimal parameters of the double-Gaussian density distribution model from the observed brightness profiles $I$ in Figure \ref{fig:bp}. The best model brightness profiles $I^\prime$ are shown as the red lines in Figure \ref{fig:bp}. We perturb the parameters ($\rho_e$, $d_{\text{front}}$, $d_{\text{sheath}}$, $\Delta L$), obtain the brightness $I^\prime(L)$, and then calculate $\chi^2$ comparing $I^\prime(L)$ with the observed brightness $I(L)$, as discussed in Sec. \ref{subsec:sd}. The various density models are obtained by the various sets of parameters which are given with the intervals of $\Delta \rho_e$ = 0.5$N_\text{Leblanc}(r)$, $\Delta d_\text{front}$ = 0.1 $R_\odot$, $\Delta d_\text{sheath}$ = 0.5 $R_\odot$ and $\Delta \Delta L$ = 0.02 $R_\odot$. The full ranges of the input parameters are shown in Figure \ref{fig:ex1}. Note that $\Delta \rho_e$ shown is normalized to the Leblanc model at the corresponding height $r$ because the electron density in the corona varies in several orders of magnitude. Once the minimum $\chi^2$ is obtained, we calculate $\chi^2$ again with the different intervals of the parameters. The full ranges of the parameters used for the second calculation are given as $\pm$ 2$\Delta^\prime$, where $\Delta^\prime$ is the initial intervals, centered on the parameters at the minimum $\chi^2$. The intervals of the parameters are $\Delta \rho_e$ = 0.1$N_\text{Leblanc}(r)$, $\Delta d_\text{front}$ = 0.02 $R_\odot$, $\Delta d_\text{sheath}$ = 0.1 $R_\odot$, and $\Delta \Delta L$ = 0.02 $R_\odot$. Not only the minimum $\chi^2$ values, we check the resultant profiles visually as shown in Figure \ref{fig:bp}.

The dependency of $\chi^2$ on the four parameters $\rho_e$, $d_\text{sheath}$, $d_\text{front}$, and $\Delta L$ is shown in Figure \ref{fig:ex1}. The top and bottom panels are $\chi^2$ values for Cut A and Cut B, respectively. Since $\chi^2$ is normalized by the model brightness $I^\prime$, the minimum $\chi^2$ value is less than unity (see color bars). As shown in Figure \ref{fig:sm}(a), the value and the shape of $I^\prime$ are sensitive to $\rho_e$. Because of this, as shown in Figures \ref{fig:ex1}(a)-(c) and \ref{fig:ex1}(e)-(g), $\chi^2$ converges at a $\rho_e$. It is also shown that $d_\text{front}$ = 0 (panels (b), (d), (f) and (h)), being consistent with the density profile of shock waves. It may imply that the shock thickness is very small and thus cannot be resolved in our observation and method. It is evident that $d_\text{sheath}$ (panels (a) and (e)) and $\Delta L$ (panels (c) and (g)) do not affect the measure of $\rho_e$. The low $\chi^2$ values (black colors) are aligned along the best $\rho_e$ value (see the vertical dashed lines in these panels). It demonstrates that our $\chi^2$ approach can provide the reasonable measure of $\rho_e$.

We repeat the $\chi^2$ fitting at all position angles. Since it is applied also in the time series, the temporal evolution of the sheath structures can be investigated. Figure \ref{fig:evol} shows the time evolution of the sheath propagating at the position angle $\zeta$ = 80$^\circ$ (Cut A). The position angle in three different images is shown in the upper panels of Figure \ref{fig:obs}. The background density in this figure has been determined independently in Section \ref{subsec:dc}. As the shock propagates outward, $\rho_e$ falls off.

Figure \ref{fig:den} shows the derived $\rho_e$ for the 2011 March 7 and 2014 February 25 CMEs. The upper panels show  $\rho_e$ versus $\zeta$ for three consecutive times (marked by the different colors). Because of the shock--streamer interaction at 20:24 UT on 2011 March 7 (see the arrow in Figure \ref{fig:obs}), we do not attempt to derive $\rho_e$ beyond $\zeta$ $\sim$ 50$^\circ$. As expected, $\rho_e$ decreases with time, and tends to  maximum around the shock leading edge ($\zeta$ $\sim$ 0$^\circ$). In the lower panels, we plot $\rho_e$ versus heliocentric distance for all position angles. $\rho_e$ decreases with height but remains high at the leading edge (marked by the downward arrows). This is a clear indication, at least to us, that the super-magnetosonic CME keeps driving shocks and it results in the higher density jump than the lateral parts that are likely decoupled from the driver and turning into a freely-propagating shock wave \citep{K2017}. \citet{K2017} showed that the speed of the shocks ahead of the CME noses is well correlated with the CME nose speeds while the shock speed in the far-flanks is not. The far-flank speeds tend rather towards the local fast magnetosonic speed. For comparison, the empirical density models of \citet{S1977} and \citet{L1998} are  shown by the dashed and solid lines, respectively.

The errors in $\rho_e$ are shown by error bars in Figure \ref{fig:den}. 
To determine the errors, we take into account the two sources of the error in $\rho_e$ as described in Section \ref{sec:err}. First, the error can arise from the uncertainty of the excess brightness. Assuming that $\rho_e$ varies slowly along the shock surface in the real corona, the fluctuations in $\rho_e$ determined within a small range of $\zeta$ could be due to the uncertainty. We have repeated our fits for every 1$^\circ$ intervals of $\zeta$ independently, and we take the average and standard deviation over $\zeta$ spanning 7$^\circ$. The other source of error is the 3D geometry of the shock modeled with the ellipsoid. The error of the ellipsoid model has been determined as shown in the Appendix of \citet{K2014}. The error is $\sim$ 4\% for all images we have analyzed. The two additional ellipsoid models considering $\pm$ 4\% error are obtained, and their projections on the images are shown as dashed lines in Figure \ref{fig:obs}. We repeat the same $\chi^2$ calculation with these ellipsoid models, and the results are considered as the error in $\rho_e$ due to the 3D geometry of the shock. Once the two errors are calculated, we use the larger one for the error in $\rho_e$.

\subsection{Density Compression Ratio} \label{subsec:dc}
The  density compression ratio, $X$ = $\rho_d$/$\rho_u$, where $\rho_d$ = $\rho_e$ + $\rho_u$, can only be determined if $\rho_u$ is known. 

We determine $\rho_u$ using the polarized brightness images prior to each event as described in Section \ref{subsec:up}. In practice, we have used the polarized brightness images observed from 00:08 UT to 14:08 UT on 2011 March 7 and from 19:08 UT on 2014 February 24 to 00:08 UT on 25. The standard polarized brightness inversion method is applied to the averaged polarized brightness images. Figure \ref{fig:up} shows the determined $\rho_u$ along the shock fronts where $\rho_e$ has been determined. Given the inhomogeneity of the background corona, the derived $\rho_u$ varies considerably along the shock fronts resulting in the `U-shaped' curves. The errors in $\rho_u$ are determined from the errors in $r$ due to the uncertainty of the 3D geometry of the shock. The errors in $r$ and $\rho_u$ are shown by the horizontal and vertical error bars, respectively. Because of the uncertainty in the background electron density \citep[e.g.,][]{hayes2001}, we also employ an empirical coronal electron density model of \citet{L1998} that provides the lowest values among the other empirical models (solid lines in Figure \ref{fig:up}). In general, measured $\rho_u$ is greater than the Leblanc model probably due to F-corona polarized component, but a portion of the 2011 March 7 shock propagates in the northern polar coronal hole in the sky planes, and the measured $\rho_u$ in this region is lower than that of the Leblanc model due to the calibration \citep[e.g.,][]{T2010}. The measure of electron density in coronal holes is even more uncertain \citep[e.g.,][]{hayes2001}. We also consider the error in $r$ when taking $\rho_u$ from the empirical model. We use the full ranges covering the observations and empirical model as well as their errors, as the upstream electron densities $\rho_u$.

Figure \ref{fig:X} shows the ratios between the downstream and upstream electron densities, i.e., compression ratios $X$. The speeds are also shown in colored lines for the fronts where $X$ is determined. The error bars are the propagation errors given by the errors in $\rho_e$ and $\rho_u$. The compression ratio peaks around the shock leading edge, as also seen in $\rho_e$ in Figure \ref{fig:den}. Note that there is no dramatic increase in $X$ around the shock leading edge of the 2014 February 25 shock as the 2011 March 7 event. It may be because of the upstream electron density $\rho_u$ and the proximity of the shock front to the CME. As seen in Figure \ref{fig:up}, $\rho_u$ at the leading edge of the February event is lower than that of the March event. The 3D angular distances between the CME noses and the the projected shock leading edges are $\sim$ 40$^\circ$ (2011 March 7) and $\sim$ 50$^\circ$ (2014 February 25). In addition, $X$ seems to be correlated with speed. However, it is interesting to note that $X$ is nearly constant in time \citep[cf.][]{B2011}, while the speed decreases with time, as clearly seen in the bottom panels. Figure \ref{fig:evol} indicates that this is due to the decrease in $\rho_u$ with height (see also Figure \ref{fig:up}). As the shocks propagate outward, $\rho_e$ is falling off together with $\rho_u$. Table \ref{tab} shows the full ranges of the determined $X$ with the averages. 

\begin{figure}
\begin{center}
\includegraphics[width=.9\textwidth]{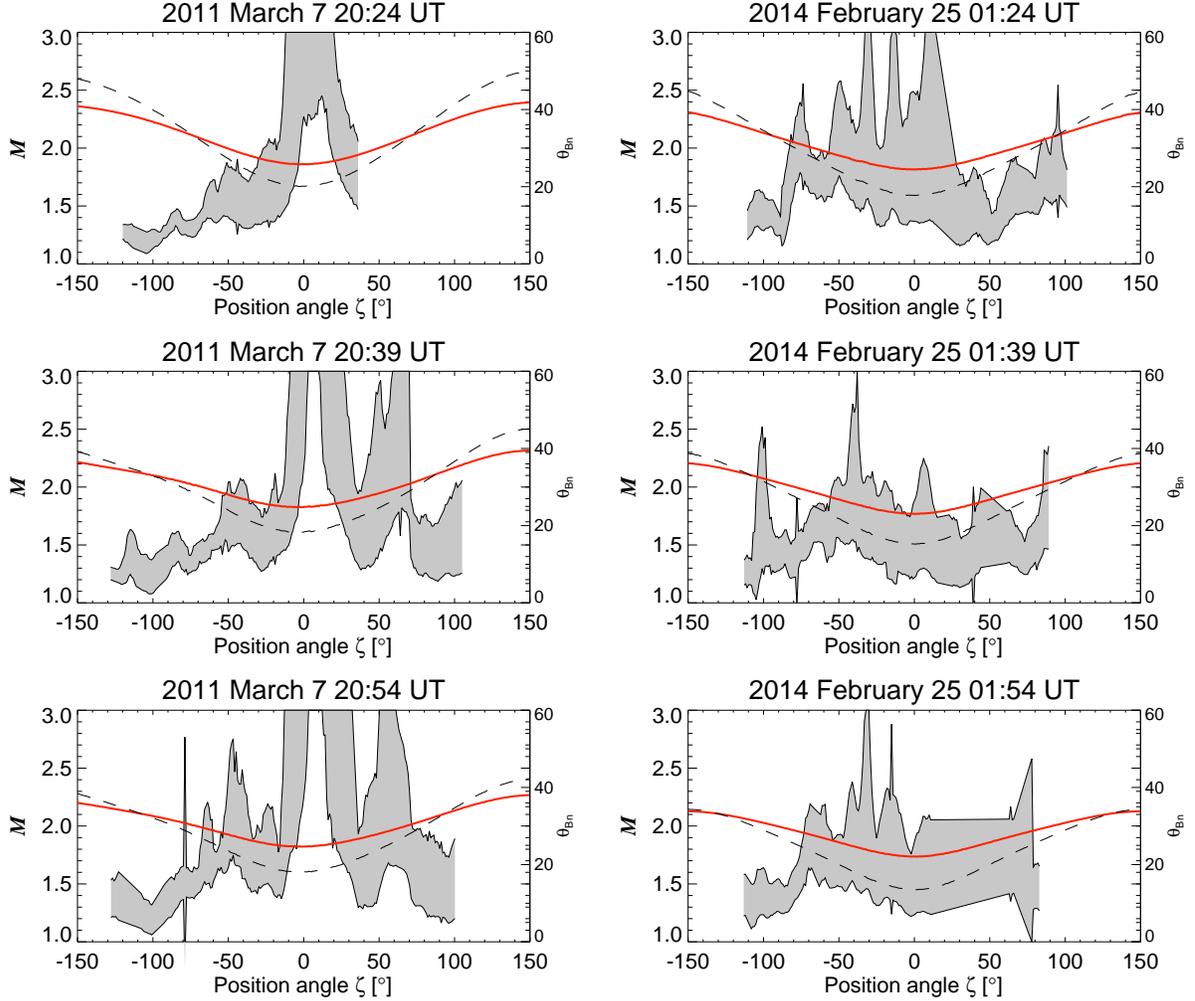}
\caption{Mach numbers $M_A$ (shaded regions) derived from the full ranges of $X$ in Figure \ref{fig:X}. Red lines indicate the critical Mach number for the determined $\theta_{Bn}$. $\theta_{Bn}$ is also plotted by dashed lines (see the $y$-axes on the right-handed side). } \label{fig:mach}
\end{center}
\end{figure}

The ratios $X$ in Figure \ref{fig:X} are found to be larger than 4 around the shock leading edge. The upper limit of $X$ is given as $(\gamma+1)/(\gamma-1)$, where $\gamma$ is the adiabatic index, and the upper limit is 4 if $\gamma$ = 5/3. It might imply that the adiabatic index $\gamma$ could be close to 1 in the low corona \citep{VD2011}. If $\gamma$ = 4/3, for example, the upper limit is 7. Alternatively, the regions where $X$ $>$ 4 are only the small parts of the shock fronts as shown in Figure \ref{fig:X}. The averages and standard deviations shown in Table \ref{tab} are lower than 4. It may imply that it is due to the errors in $\rho_e$ and/or $\rho_u$, and the actual $X$ is less then 4. As discussed in Section \ref{sec:de}, the brightness profiles taken close to the shock leading edge could be slightly contaminated by the pileup plasma of the following CME.

Several attempts have been made in the past literature to determine the density ratios $X$ from white light \citep{O2009,B2010,B2011} and EUV \citep{Kv2011,koulou2014,L2015}, and radio observations \citep{Ma2011}. \citet{O2009} determined $X$, ranging between 1.2 and 2.8 for 11 CMEs that were faster than 1500 km s$^{-1}$ observed by LASCO C2 coronagraph. \citet{B2010, B2011} also used the LASCO C2 images and found that $X$ $\approx$ 1.2--3.0. Although these values were based on projected brightness profiles or single-view reconstructions, they are within the range of our values determined via more sophisticated 3D reconstructions. Similarly, for the variation along the shock front, \citet{B2011} concluded that the compression ratio varied from 1.2 at the lateral flank to 3.0 at the shock leading edge. Our results validate past efforts and show that the compression ratio, in the middle corona at least, generally ranges from 2 to 4 (see Table \ref{tab}).
In addition, compression ratios were also determined in the early stage of events from EUV \citep{Kv2011,koulou2014,L2015} and radio \citep{Ma2011,koulou2014} observations.  These compression ratios are slightly larger than unity and tend to be smaller than those derived from white light observations. Note that these measurements have been done close to the Sun, indicating the early stage of the events. It may imply that the shock is still developing in those heights.

While the derived $X$ values are consistent with past results, their temporal evolution differs in that the values are more or less constant in contrast to the results in \citet{B2010,B2011}. \citet{B2011} showed that the peak $X$ value declines from $\sim$ 3.0 to $\sim$ 1.5 as the shock leading edge travels from $\sim$ 3 $R_\odot$ to $\sim$ 6 $R_\odot$.  Note that our CMEs are very fast, and the speeds at the times of the latest images are still over 1500 km\,s$^{-1}$. \citet{K2017} showed that the speeds of these driver CMEs are much faster than the local fast-mode wave speeds and are able to generate shocks during our measurements. The fast speeds may be responsible for the high $X$ even in higher altitudes.

\subsection{Alfv\'{e}nic Mach Number, $M_A$}
To relate our results to modeling effort and past literature, we proceed to compute the Alfv\'{e}nic Mach number, $M_A$, under the assumption of $\gamma$ = 5/3. $M_A$ can be estimated from the compression ratio, $X$ \citep[e.g.,][and references therein]{B2011}. $M_A$ is a function of plasma $\beta$ and $\theta_{Bn}$, and we could simplify to the case of $\beta$ $\rightarrow$ 0 for the coronal medium \citep{K2013b}. $\theta_{Bn}$ is the angle between the magnetic fields and the shock. We assume that the coronal magnetic fields are purely radial in our height range and determine $\theta_{Bn}$ using the ray-like trajectories of the shocks shown in \citet{K2017}. The Mach number of an oblique shock is given by \citep[e.g.,][and references therein]{B2011}
\begin{equation}
M_{A}^2 = (M_{A\bot}\sin{\theta_{Bn}})^2+(M_{A\parallel}\cos{\theta_{Bn}})^2 \, ,
\end{equation}
where $M_{A\bot}$ = $[0.5X(X+5)/(4-X)]^{1/2}$ and $M_{A\parallel}$ = $X^{1/2}$, assuming the adiabatic $\gamma$ = 5/3.

\begin{figure}
\begin{center}
\includegraphics[width=.5\textwidth]{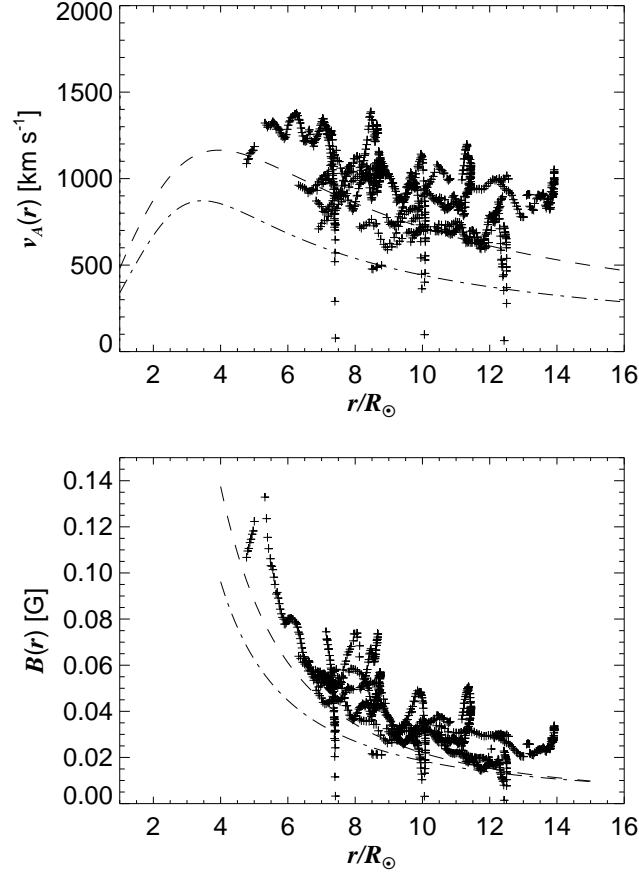}
\caption{Upper panel: Local Alfv\'{e}n speed, $v_A$ (plus symbols), inferred from the Mach numbers in Figure~\ref{fig:mach}. Dashed and dash-dotted lines refer to $v_A$ that are modeled with the empirical electron density models of \citet{L1998} and \citet{S1977}, respectively, together with the empirical magnetic field function of $B$ = 2.2$(r/R_\odot)^{-1}$. Lower panel: Magnetic field strength $B$ inferred from $v_A$ in the upper panel together with the determined $\rho_u$. Lines in dashed and dash-dotted are radial magnetic field profiles given in \citet{M1999} above the quiet Sun and \citet{DM1978} above an active region, respectively.} \label{fig:local}
\end{center}
\end{figure}

Figure \ref{fig:mach} shows the derived $M_A$ from the full ranges of $X$ in Figure \ref{fig:X}. Dashed lines are the estimated $\theta_{Bn}$. Because of the assumption of purely radial magnetic fields, $\theta_{Bn}$ decreases monotonically with time. The full ranges of $M_A$ and the averages are given in Table \ref{tab}. Similarly to $X$, $M_A$ seems to be more or less constant in time, which is contrary to the prevailing wisdom \citep[as shown in][]{B2011}. This does not seem unreasonable since the local Alfv\'{e}n speed should decrease with height. As a shock propagates outward the shock speed should also fall, and it results in a (relatively) constant $M_A$ since $M_A$ = $V_\text{sh}$/$v_A$ (Figure \ref{fig:local}).

Since $M_A$ = $V_\text{sh}$/$v_A$, we can use the Mach number to estimate the Alfv\'{e}nic speed across the shock front. To that end, we need to move to the shock frame by estimating the ambient solar wind speed, $V_{SW}$. Then, the true shock speed $V_{sh}$ = $V^\prime_\text{sh}$-$V_\text{SW}$$\cdot$$\cos{\theta_{Bn}}$, where $V^\prime_\text{sh}$ is the measured one. Since $V_{SW}$ measurements does not exist in these coronal heights, we turn to a formula given in \citet{S1997},
\begin{equation}
V_{SW}^2=2A(r-r_1) \, ,
\end{equation}
where $A$ = 3.6 m s$^{-1}$ and $r_1$ = 2.1 $R_\odot$. In this manner, $v_A$ can be estimated and is shown in the upper panel of Figure \ref{fig:local}. The upstream $v_A$ decreases with height and time but our estimates are higher than those predicted by empirical models of density and magnetic field (see the dashed and dash-dotted lines). This may {\it not} be unreasonable since our two case studies are very fast CMEs with speeds higher than 2000 km s$^{-1}$. Note that the magnetic field strengths inferred from the determined $v_A$ and $\rho_u$ in the lower panel of Figure \ref{fig:local} are consistent with the empirical models given in \citet{M1999} and \citet{DM1978}. We need further investigations including slower CMEs to see the general characteristics of shocks associated with CMEs.

In addition, plasma $\beta$ can also be determined from the estimated upstream $v_A$ by a relation that $\beta$ = ($2$/$\gamma$)($c_S$/$v_A$)$^2$, if sound speed $c_S$ is known \citep[e.g.,][]{K2013b}. We obtain that $\beta$ = 0.06 $\pm$ 0.02 in our height range, assuming $\gamma$ = 5/3 and $c_S$ = 200 km s$^{-1}$. It is consistent with the low $\beta$ assumption for the corona widely used in this field.

\subsection{Physical Implications for SEP Acceleration}
One of the challenges in SEP research is to explain the wide longitudinal distribution of SEPs originating in a single flare and CME event \citep[][ and references therein]{D2016}. The large widths of SEP-associated CMEs are the obvious candidate. While the spatial and temporal relationships between the CME-driven shock wave and SEP injections seem to support this assertion \citep[e.g.,][]{R2012,R2016,La2014,La2016}, it remains unclear whether the shocks are capable of accelerating particles at these locations since the properties of the local plasma environment (i.e., seed particle populations, turbulence levels) in the coronal heights are largely unknown. However, we have presented a method that allows us to derive the electron density distribution (in 3D) in the shock and hence obtain the density compression ratio and consequently the Mach number, Alfv\'{e}n speed and other parameters (under further assumptions). So we can, at least, deduce whether, and most importantly where, our shocks have the potential to accelerate particles by comparing our Mach numbers to the critical Mach number, $M_c$, \citep[e.g.,][]{B2011}.

Red lines in Figure \ref{fig:mach} show the critical Mach number corresponding to the estimated $\theta_{Bn}$. The critical Mach number $M_c$ for a collisionless shock is taken from Figure 2 in \citet{Tr2009}, for the case of $\beta$ $\approx$ 0 (we have obtained $\beta$ $<$ 0.1). It seems that our shocks are supercritical ($M_A$ $>$ $M_c$) over significant portion of their extent. We also see that the condition extends for longer periods than those reported by \citet{B2011}. Note that our case events are associated with longitudinally-wide SEP events \citep{P2013,Ri2014,La2016}. Our work here provides significant additional support to these previous works that the SEPs are produced at the CME shocks and that the wide extent of these shocks is the reason for the distribution of SEPs over a very wide range of heliospheric longitudes.

\section{SUMMARY AND CONCLUSION} \label{sec:con}
We present a new method that, using multi-viewpoint observations and forward modeling techniques, enables us to model the observed brightness of shock fronts in white light coronagraphic observations and extract the three-dimensional electron density distribution across the fronts. We apply the method to two case studies; the CME events on 2011 March 7 and 2014 February 25. Both CMEs were fast ($>$ 2000 km s$^{-1}$) and  wide ($\geq$ 200$^\circ$) halo events. The 3D reconstructions of the CMEs and their shock envelopes were reported  in a separate study \citep{K2017}. The shock envelope is based on an ellipsoidal fitting and its white light emission is assumed to arise from electrons distributed in a thin shell over the surface of the shock. The electron distribution is assumed to be a double-Gaussian described by three free parameters. By varying the parameters and integrating the resulting density profiles along the known (from the ellipsoid model) LOS, we can derive modeled brightness profiles to be compared against the observed ones. We locate the best set of free parameters via a $\chi^2$ minimization approach. We use two methods to estimate the background density; the empirical \citet{L1998} model and the standard polarized brightness inversion technique using the pre-event polarized brightness images. In this way, we obtain an upper and lower estimate for the background density and thus we can bound the resulting density compression ratio. The final results of this exercise are the excess electron density $\rho_e$, density ratio $X$ and Alfv\'{e}n Mach number $M_A$ across the full shock front. This is the first time that the properties of white light shocks are quantified in this way. Our results corroborate past estimates, mostly based on single viewpoint observations \citep{O2009,B2010,B2011} and are fully consistent with the presence of radio emissions and SEPs in the two cases.

Our findings can be summarized as follows,
\begin{enumerate}
\item The density excess, $\rho_e$ peaks at and around the CME leading edge and the peak is maintained during the time interval we analyzed (Figure \ref{fig:den}). This finding indicates that the two CMEs drive bow-type shock waves around their noses while the shock waves in the lateral flank propagate nearly freely, (i.e. as blast waves). This is the same conclusion we reached in the previous paper \citep{K2017} using the 3D speeds and a simple model.
\item The density compression ratio, $X$ (Figure \ref{fig:X}), and Alfv\'{e}nic Mach number, $M_A$ (Figure \ref{fig:mach}), also vary with the position angle $\zeta$, i.e., their maximum occurs around the CME nose and correlates with the shock speed. Both $X$ and $M_A$ remain relatively constant in time and distance despite the decrease of the shock speed (see also Table \ref{tab}). It is not unreasonable because the local Alfv\'{e}n speed $v_A$ also decreases as the shocks propagate further out (Figure \ref{fig:local}).
\item The shocks are supercritical over a wider spatial range, and they last longer than those of what has been reported by \citet{B2011}.
\item The averages of $X$ and  $M_A$ are 2.1--2.6 and 1.1--1.8, respectively (see Table \ref{tab}). 
\end{enumerate}

Once again, we show (this time via density analysis) that the diffuse white light emission ahead of fast CMEs outlines the shock sheath. Our 3D analysis suggests that the density compression is sufficiently high for the production of high energy particles and that the conditions last for, at least, tens of minutes over a wide range of position angles around the CME nose. It suggests that the CME-associated shocks can account both for the SEP production and their longitudinal distribution.

Our method builds upon and greatly extends past work \citep{Tn2006,O2009} by using the 3D reconstruction information of the shock envelope to deconvolve the density distribution from projection effects and to derive the 3D speed, upstream and downstream densities, Mach number and other parameters across the full shock front. The method holds promise for improving the extraction of quantitative information from shocks in the corona using remote sensing observations. It can readily use observations from different vantage points, including from the imagers \citep{H2013,V2016} aboard the upcoming Solar Orbiter \citep{Mueller2013} and Solar Probe Plus \citep{Fox2016} missions to be compared with direct in situ measurements from these missions. The technique will also greatly improve the results from any future joint coronagraphic imaging and off-limb spectroscopy of shocks \citep{B2011, voubempo2012}. We plan to test the double-Gaussian density profile along the shock normal, 3D geometry and the $\chi^2$ approach with numerical simulations. Further analyses, including slower CME--shock events, will lead to a better understanding of the shock properties and its relation to SEPs and other phenomena such as EUV waves and radio bursts.

\begin{table}
\caption{The determined density compression ratios $X$ and Mach numbers $M_A$ with the corresponding ranges of height $r$. This table provides the full ranges of these values with the averages and standard deviations.}  \label{tab}           
\label{table:1}      
\centering                          
\begin{tabular}{c c c c c c}        
\hline\hline                 
Date & Time & $r$/$R_\odot$ & $X$ & $M_A$ \\
\hline                        
   2011 March 7 & 20:24 UT & 4.8--7.4       & 1.4--4.6 (2.4 $\pm$ 0.8) & 1.2--4.2 (1.8 $\pm$ 0.6) \\
                & 20:39 UT & 6.9--10.1      & 1.4--4.3 (2.4 $\pm$ 0.7) & 1.2--4.6 (1.7 $\pm$ 0.5) \\
                & 20:54 UT & 8.5--12.5      & 1.3--4.8 (2.6 $\pm$ 0.8) & 1.2--5.2 (1.8 $\pm$ 0.5) \\
\hline\hline
   2014 February 25 & 01:24 UT & 6.3--8.7   & 1.6--3.0 (2.3 $\pm$ 0.3) & 0.1--2.1 (1.1 $\pm$ 0.6) \\
                    & 01:39 UT & 8.1--11.4  & 1.5--2.9 (2.1 $\pm$ 0.3) & 1.3--1.9 (1.6 $\pm$ 0.1) \\
                    & 01:54 UT & 10.0--13.9 & 1.6--2.9 (2.3 $\pm$ 0.4) & 1.3--1.9 (1.6 $\pm$ 0.2) \\
\hline                                   
\end{tabular}
\end{table}

\begin{acknowledgements}
We would like to thank Rob Decker and David Lario for insightful discussions. This work of R.Y. K. and A.V. is supported by  NASA Grant NNX16AG86G issued under the HSR Program. The SECCHI data are produced by an international consortium of the NRL, LMSAL and NASA GSFC (USA), RAL and Univ. Bham (UK), MPS (Germany), CSL (Belgium), IOTA and IAS (France). The editor thanks two anonymous referees for their assistance in evaluating this paper.
\end{acknowledgements}




\end{document}